\documentclass[runningheads]{llncs}

\usepackage[T1]{fontenc}
\def\doi#1{\href{https://doi.org/\detokenize{#1}}{\url{https://doi.org/\detokenize{#1}}}}
\usepackage{graphicx}

\usepackage{marvosym}
\usepackage{listings}
\begin{document}
%
\title{Cerebrovascular Segmentation via Vessel Oriented Filtering Network}
%
%
\author{Zhanqiang Guo\inst{1} \and
Yao Luan \and
Jianjiang Feng\inst{1}${\textsuperscript{(\Letter)}}$ \and Wangsheng Lu \and Yin Yin \and Guangming Yang \and Jie Zhou}
\authorrunning{Zhanqiang Guo et al.}
\institute{Department of Automation, Tsinghua University, Beijing, China}
%
%
\maketitle              
\begin{abstract}
Accurate cerebrovascular segmentation from Magnetic Resonance Angiography (MRA) and Computed Tomography Angiography (CTA) is of great significance in diagnosis and treatment of cerebrovascular pathology. Due to the complexity and topology variability of blood vessels, complete and accurate segmentation of vascular network is still a challenge. In this paper, we proposed a Vessel Oriented Filtering Network (VOF-Net) which embeds domain knowledge into the convolutional neural network. We design oriented filters for blood vessels according to vessel orientation field, which is obtained by orientation estimation network. Features extracted by oriented filtering are injected into segmentation network, so as to make use of the prior information that the blood vessels are slender and curved tubular structure. Experimental results on datasets of CTA and MRA show that the proposed method is effective for vessel segmentation, and embedding the specific vascular filter improves the segmentation performance.

\end{abstract}
\section{Introduction}
Cerebrovascular disease is one of the leading causes of death and disability worldwide \cite{weissbrod2020neurologic}. Magnetic Resonance Angiography (MRA) and Computed Tomography Angiography (CTA) are often applied to imaging the cerebrovascular system in clinical diagnosis. Accurate segmentation of cerebral vasculature from MRA and CTA images is the first step for many clinical applications. However, segmenting cerebral vessels is very challenging due to various difficulties, such as the complex cerebral vascular structure, high degree of anatomical variation, small vessel size, poor vessel contrast, and data sparseness \cite{almi2012automatic,gouw2011heterogeneity}.

In recent decades, a number of automatic (or semi-automatic) vascular segmentation methods have been proposed \cite{lesage2009review,moccia2018blood}. Existing cerebrovascular segmentation algorithms can be coarsely divided into two categories: traditional algorithms and deep learning based algorithms. The key of the first type of method is to artificially formulate some rules to extract blood vessel features, which usually take into account the fact that the vessels are thin and slender tubular structure. For example, many segmentation methods are based on features computed from hessian matrix \cite{frangi1998multiscale,jerman2016enhancement}. Combined with the features extracted by the rules formulated artificially, some other traditional methods are also used for vessel segmentation. Yang et al. \cite{yang2014geodesic} presented a geodesic activate contour model to segment blood vessels, which adaptively configures parameters. Zhao et al. \cite{zhao2014multi} proposed a likelihood model based on level set, achieving good results for segmentation of vascular network. Lv et al. \cite{lv2020parallel} proposed a parallel algorithm based on heterogeneous Markov random field to segment vessels. However, these methods depend heavily on domain knowledge which may be inaccurate, and thus are not robust to image quality.

Recently, deep learning methods are widely used in the field of medical image processing. The segmentation algorithms based on convolutional neural network have been shown to produce state of the art results in various medical segmentation tasks \cite{taher2021automatic,simpson2019large,liu2021review,xia2021nested}. One of the most notable methods is 3DUnet \cite{cciccek20163d}. Similar to it, Fausto et al. \cite{milletari2016v} used convolution instead of max-pooling and residual connections to improve performance. Inspired by the inception model \cite{szegedy2017inception}, Pedro et al. \cite{sanchesa2019cerebrovascular} proposed Uception for vascular segmentation, which increases the network size to have a better representation. To reduce computational burden and memory consumption in 3D segmentation, 2D orthogonal cross-hair filters were formulated in DeepVesselNet \cite{tetteh2020deepvesselnet}. Zhang et al. \cite{zhang2020cerebrovascular} proposed RENet to remove redundant features and retain edge information in shallow features, and reported good results in segmentation of cerebral vessels. Based on Multiple-U-Net, Guo et al. \cite{guo2021cerebrovascular} proposed a novel network for vessel segmentation. Aimed for CTA images, Fu et al. \cite{fu2020rapid} proposed a framework based 3D convolutional neural network for segmenting cerebrovascular networks. However, these methods fail to explicitly consider the peculiarities of the vascular structure, namely, blood vessels are curved and elongated tubular structure. It is difficult and unpredictable to rely solely on network training to learn these characteristics of vessels from training data, especially when the size of training set is limited, which is typical in medical image analysis.

In response to the problems mentioned above, we propose a Vessel Oriented Filtering Network (VOF-Net) that combines domain knowledge and deep learning method. First, we propose a vessel orientation estimation network to calculate the orientation field of blood vessels, through which vessel oriented filters are designed. Then, we feed the extracted vessel features to a 3DUnet based vessel segmentation network, which is trained to combine information from features extracted by oriented filters. Experiments on CTA and MRA datasets show that our proposed method yields better result than other popular vascular segmentation methods.

\begin{figure}[tbp]
	\centering
	\centerline{\includegraphics[width=12.5cm]{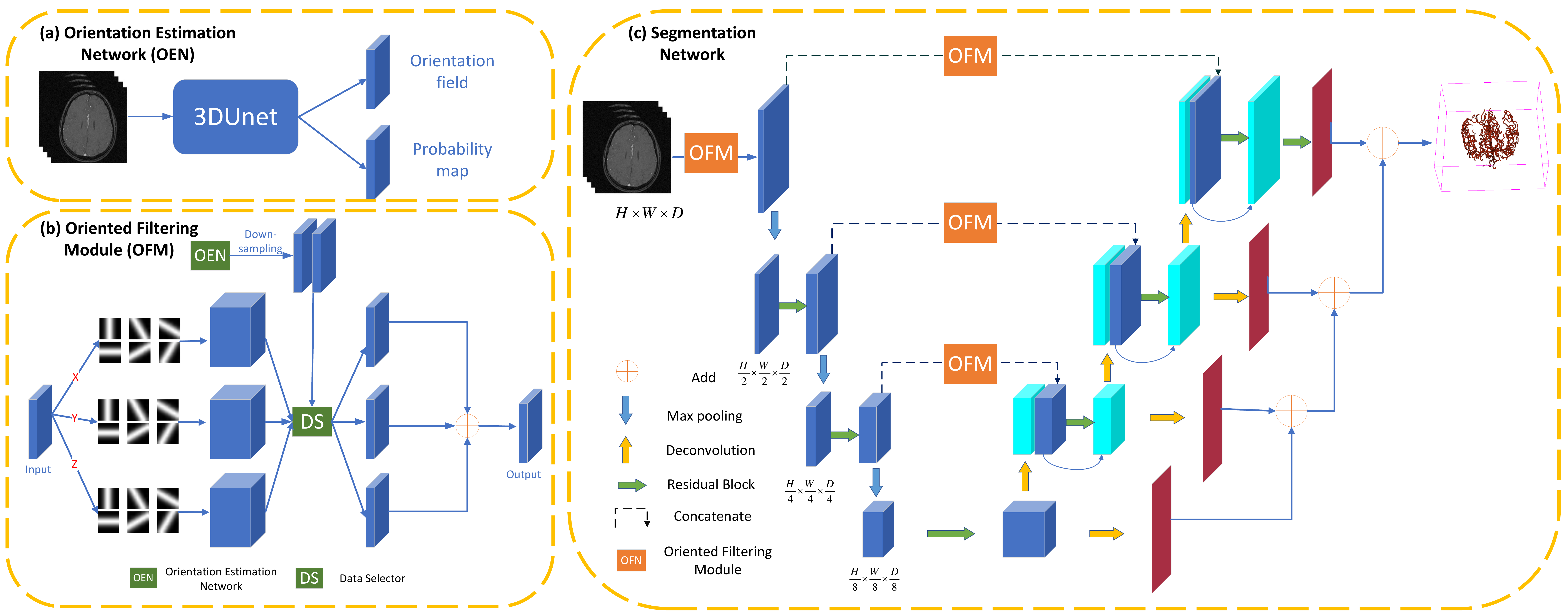}}
	\caption{Proposed VOF-Net includes Orientation Estimation Network, Oriented Filtering Module, and Segmentation Network.}
	\label{fig1}
\end{figure}

\section{Methodology}

Our proposed system, as shown in Fig.~\ref{fig1}, consists of three modules: Orientation Estimation Network, Oriented Filtering Module, and Segmentation Network.

\subsection{Orientation Estimation Network}

\begin{figure}[hbp]
	\centering
	\includegraphics[width=9.5cm]{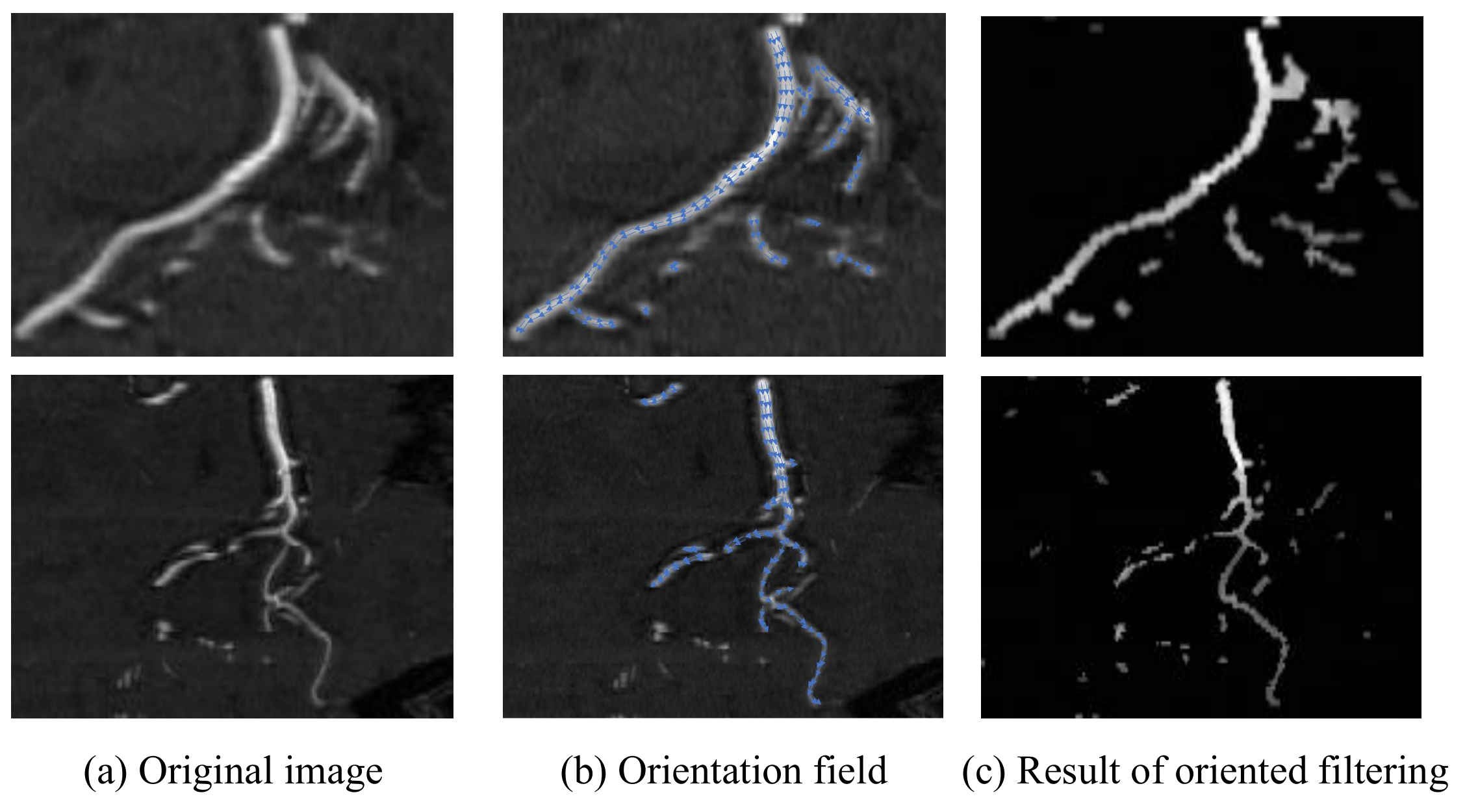}
	\caption{Orientation field (blue lines) of blood vessels and the result of oriented filtering.} 
	\label{fig2}
\end{figure}

The orientation estimation network in our framework is designed based on 3DUnet, which uses residual connection instead of convolution structure. The vessel orientation field of a training sample with ground-truth vessel segmentation is obtained in this way: First, we refine the binary image of the blood vessel to procure the centerline, of which each point is assigned the value by calculating its tangent orientation. Based on the orientation field of centerline, the orientation field on the vascular segmentation map is obtained via nearest neighbor interpolation. The calculated orientation field of a training sample is shown in Fig.~\ref{fig2}(b).

The network predicts the orientation field as well as the probability map of blood vessels. Note that this vessel probability map may not be accurate enough as the final segmentation result and just useful for determining where there may be vessels. We choose the cosine distance as the loss of orientation prediction:
\begin{equation}
	L_{{\rm cos}}=\frac{1}{|C|}\sum_{i\in C}(1 - \frac{<o_i, O_i>}{|o_i|\cdot |O_i|})
\end{equation}
where $C$ represents the set of voxels belonging to the blood vessels. $O_i, o_i$ correspond to the actual and predicted vessel orientation. $<\cdot,\cdot>$ indicates calculating the inner product of two vectors. The dice loss is used to supervise the prediction of the vessel probability map:
\begin{equation}
	L_{{\rm dice}}=1-\frac{2\sum_{i}(t_iy_i)+\varepsilon}{\sum_i(t_i^2+y_i^2)+\varepsilon}
\end{equation}
where $t_i\in\{0,1\}$ represents the label value at voxel $i$, and $y_i$ represents the predicted value of blood vessel probability output by the network. $\varepsilon$ is a small number used to avoid being equal to 0 in the denominator. The loss function is the weighted value of the above two losses:
\begin{equation}
	L_{{\rm orient}} = L_{{\rm dice}} + \mu L_{{\rm cos}}.
\end{equation}
Then, we post-process the orientation field output from OEN to obtain 2-D quantized orientation field. Specifically, we project the calculated orientation at a voxel on the three 2-D orthogonal planes (coronal plane, sagittal plane, transverse plane) and quantify orientation into 6 categories. Take the coronal plane as an example:
\begin{equation}
	d_{i}=\lfloor \frac{O_i}{30^\circ} \rfloor + 1
\end{equation}
where $O_i (0^\circ \leq O_i < 180^\circ)$ is the projection direction of coronal plane at location $i$.

\subsection{Oriented Filtering Module}

\begin{figure}[htbp]
	\centering
	\centerline{\includegraphics[width=7.5cm]{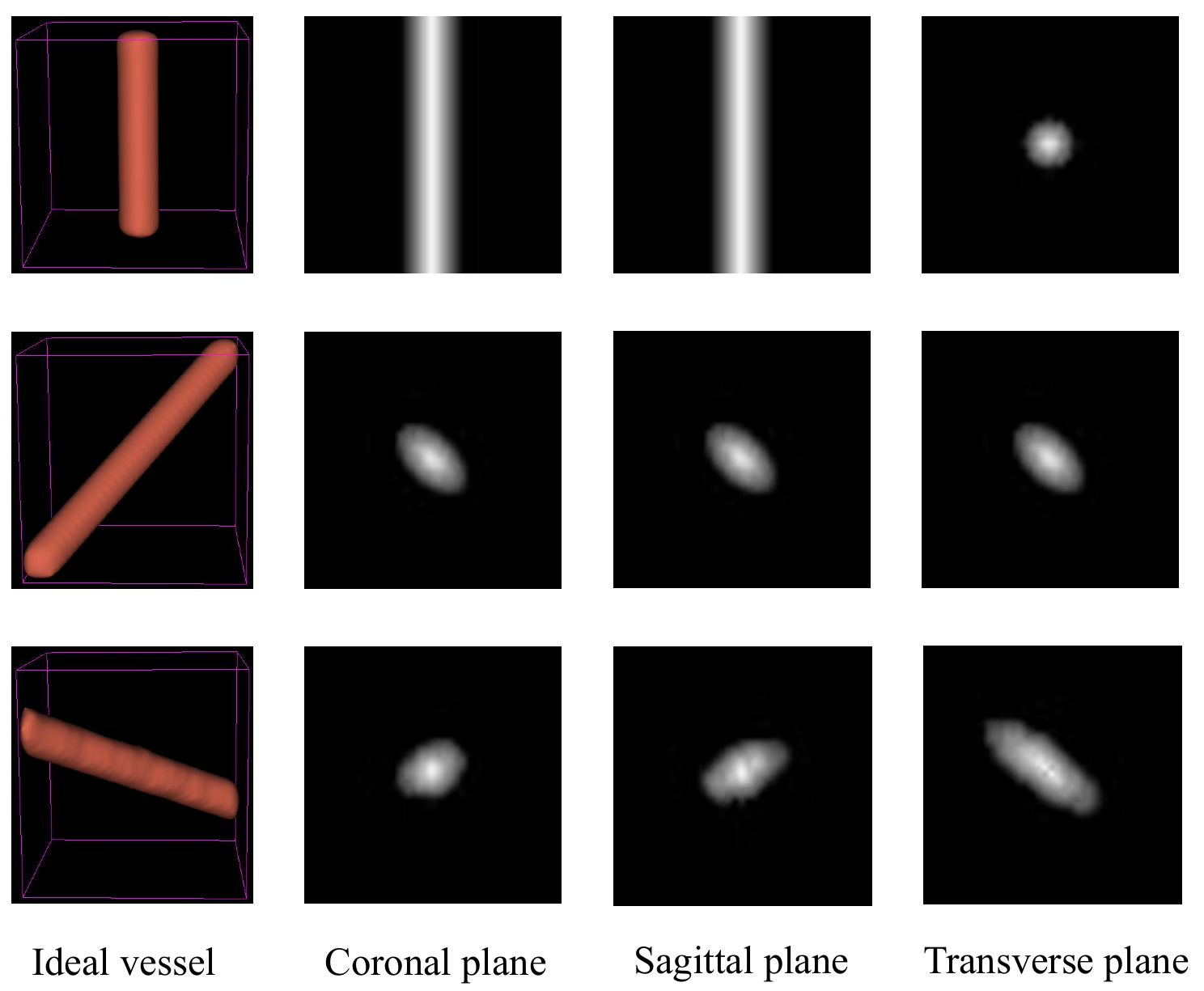}}
	\caption{Three ideal blood vessels with different orientations and their corresponding cross sections in the center.}
	\label{fig3}
\end{figure}

The oriented filtering module is a module of the whole segmentation network where the input is the deep feature map of segmentation network and the output is the result of oriented filtering. To reduce computational burden, we divide the feature map into three orthogonal planes, in which the convolution operation is performed with six Gaussian rectangular boxes and six Gaussian ellipsoid boxes, because the section of blood vessel is rectangular or elliptical (shown in Fig.~\ref{fig3}). Then we select the results of oriented filtering according to the vessel orientation field and vessel probability obtained in the previous step, which are kept the same size as the feature map by downsampling. Take one of the sections as an example, the new feature at location $i$ is computed as:
\begin{equation}
	f_i'=I(p_i\leq \delta)f_i+I(p_i>\delta)\sum_{k=1}^{6}I(d_i=k)[R_k(f_i)+E_k(f_i)]
\end{equation}
where $f_i,f_i'$ correspond to the values before and after the feature updating. $p_i$ is the predicted probability value of the blood vessels, and $d_i$ is the predicted quantization orientation in this plane. $R_k$ and $E_k$ respectively indicate that there is an oriented filtering operation with the rectangular box or ellipsoidal box, which is achieved by artificially designing convolution kernels. $I(\cdot)$ is the indicative function. 

This formula indicates that when the predicted probability value of the blood vessel is greater than a certain threshold, the result of filtering is selected according to predicted orientation, otherwise the feature value is kept unchanged. The effect of this process is to select features obtained by operating convolution with different direction kennels, represented by Data Selector in Fig.~\ref{fig1}. Finally, the calculated values of the features on the three cross-sections are added as the new feature value. The output of oriented filtering module is shown in Fig.~\ref{fig2}(c), if the input is the original image. 
\subsection{Segmentation Network}
The segmentation network is customized based on 3DUnet. To make full use of the prior knowledge of vessel structure, Oriented Filtering Module is added at the crossing junction between the contraction path and expansion path. The input image is also processed by this module because fine vascular structure features can be extracted from the original image. Furthermore, the output of each layer is deconvoluted to the original image size during up-sampling (the red feature block in Fig.~\ref{fig1}), and then superimposed upwards in turn, so that the deep features are supervised. The loss is also added at each layer:
\begin{equation}
	L_{{\rm seg}}=\lambda_1L_{{\rm seg}}^{(1)}+\lambda_2L_{{\rm seg}}^{(2)}+\lambda_3L_{{\rm seg}}^{(3)}+\lambda_4L_{{\rm seg}}^{(4)}
\end{equation}
where $L_{{\rm seg}}^{(i)}$ represents the dice loss of $i$-th layer at up-sampling and the groundtruth at $i$-th layer is obtained by  down-sampling the label.

\section{EXPERIMENTAL RESULTS}

\subsection{Data and Implementation Details}

Our VOF-Net was implemented on a PyTorch framework. The hyperparameter $\mu$ was set to 1 and $\delta$ was set to 0.1. $\lambda_1 \sim \lambda_4$ were 1.0, 0.8, 0.4, 0.2, respectively. In the two stages of training (Orientation Estimation Network and Segmentation Network), adaptive moment estimation (Adam) was employed. In addition, a poly learning rate policy \cite{zhao2017pyramid} was used, with initial learning rate 0.001 and power 0.9. The batch size was set to 4 during training due to the limitations of GPU memory.

We evaluated our proposed method on two datasets. The first dataset is the public dataset\footnote{https://public.kitware.com/Wiki/TubeTK/Data} consisting of 42 3D time-of-flight MRA volumes with labeled vessels, with voxel spacing of $0.5\times0.5\times0.8$ mm$^3$ and volume size of $448\times448\times128$ voxels.  The second dataset includes CTA images of 31 patients, provided by a local hospital. The voxel spacing of the CTA images is  $0.5\times0.5\times0.6$ mm$^3$ and the volume size is cropped to $320\times320\times208$, leaving the middle cerebrovascular area. In experiment, patches of size $64\times64\times64$ cropped from the MRA or CTA volume are fed into the network. We evaluated the results based on the following metrics: Dice similarity coefficient (DSC); Sensitivity (Sen). In addition, Average Hausdorffs Distance (AHD) was calculated. AHD between two point sets $P$ (segmentation prediction) and $L$ (segmentation label) is defined as:
\begin{equation}
	d_{{\rm AHD}}(P,L) = \frac{1}{2}(\frac{1}{|P|}\sum_{x\in P}\min_{y\in L}d(x,y) + \frac{1}{|L|}\sum_{y\in L}\min_{x\in P}d(x,y))
\end{equation}
where $d(\cdot, \cdot)$ is the distance of two points. Compared to DSC, AHD has the advantage that it takes voxel localisation into consideration, which is important for vessel segmentation \cite{taha2015metrics}.

\subsection{Results and Comparison}

\begin{figure*}[h]
	\centering
	\centerline{\includegraphics[width=11cm]{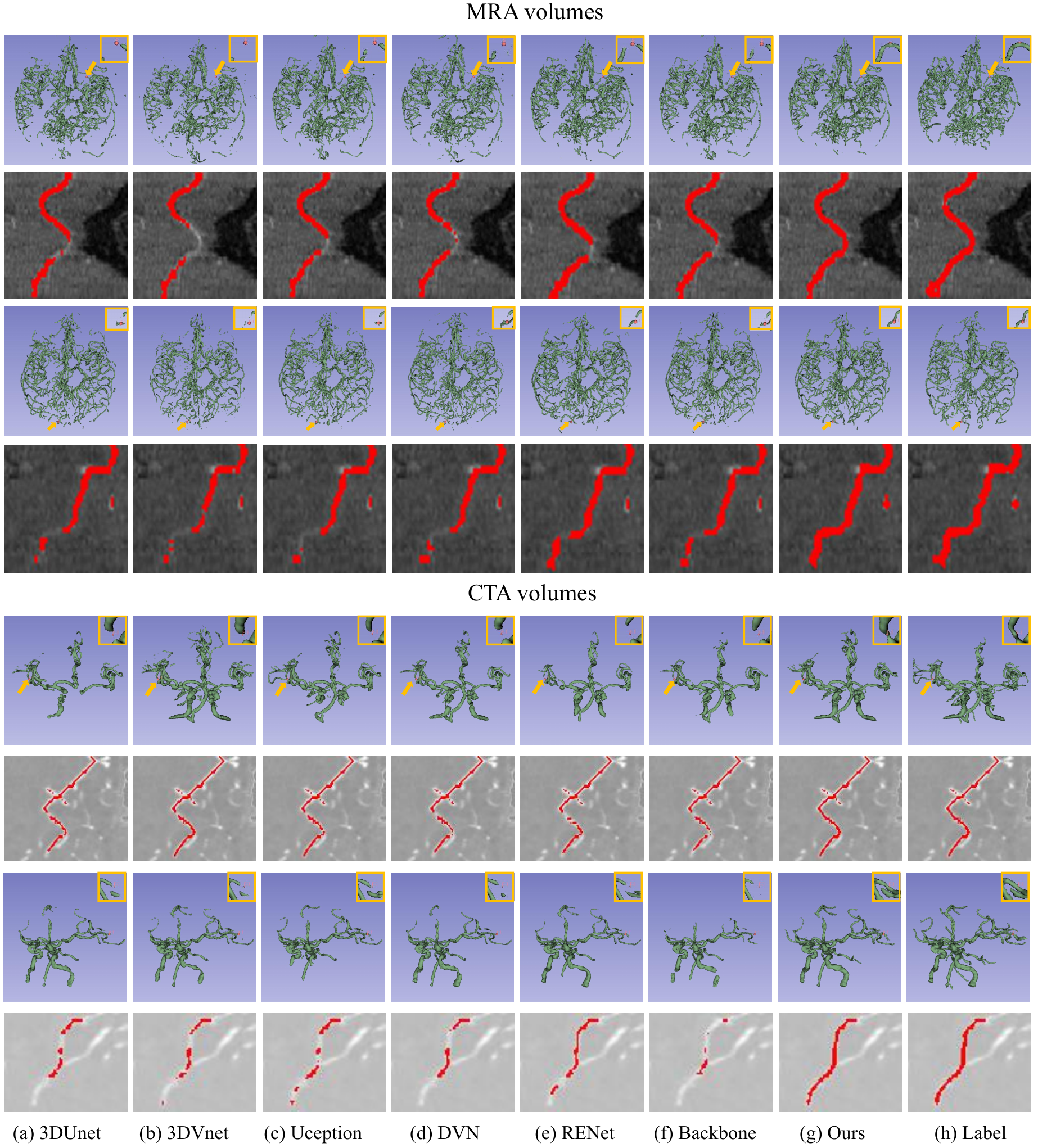}}
	\caption{Illustration of cerebrovascular segmentation results of different methods on four volumes (two MRA images and two CTA images).}
	\label{fig4}
\end{figure*}

We evaluate the performance of the orientation field estimation module with cosine distance between predicted and labeled ones. For the points on the centerlines, the average cosine distance is 0.092 (that is 24$^{\circ}$) on MRA dataset and 0.124 (that is 28$^{\circ}$) on CTA dataset, indicating that the calculated orientation field is reliable.
\begin{table}[h]
	\centering
	\small
	\caption{Comparison with other methods on two datasets.}
	\label{tab1}
	\begin{tabular}{ c | c | c | c | c | c | c | c }
		\hline
		 & \multicolumn{3}{|c|}{\textbf{MRA dataset}} & \multicolumn{3}{|c|}{\textbf{CTA dataset}} &  \\
		\hline
		\textbf{Method}             & \textbf{Dice(\%)} & \textbf{Sen(\%)}       & \textbf{AHD(mm)} & \textbf{Dice(\%)} & \textbf{Sen(\%)}       & \textbf{AHD(mm)} & \textbf{Params} \\
		\hline
		3DUnet \cite{cciccek20163d}    &  67.14 & 64.42 & 0.942  & 68.13 & 62.52 & 1.349 & 2.10M \\
		3DVnet \cite{milletari2016v}   &  59.85 & 56.13 & 1.044 & 72.54 & 73.32 & 1.062 & 4.13M \\
		Ucep \cite{sanchesa2019cerebrovascular} &  65.51 & 62.62 & 1.017 & 70.86 & 69.03 & 1.246 & 8.62M \\
		DVN \cite{tetteh2020deepvesselnet}      &  64.11 & 60.25 & 1.016 & 74.49 & 73.46 & 1.075 & 4.69M \\
		RENet  \cite{zhang2020cerebrovascular}  &  68.54 & 67.29 & 0.851 & 75.01 & 74.12 & 1.137 & 39.28M \\
		\hline
		Backbone &  66.97 & 64.55 & 0.871 & 69.23 & 62.48 & 1.352 & \textbf{2.00M} \\
		Ours     &  \textbf{69.08} & \textbf{69.74} & \textbf{0.828} & \textbf{75.20} & \textbf{78.08} & \textbf{0.843} & 4.31M\\
		\hline
	\end{tabular}
\end{table}

To evaluate the effectiveness of our proposed segmentation algorithm, we reproduced two popular 3D medical image segmentation networks (3DUnet \cite{cciccek20163d}, 3DVnet \cite{milletari2016v}). Furthermore, we also reproduced three state-of-the-art algorithms for vessel segmentation, including DeepVesselNet (DVN) \cite{tetteh2020deepvesselnet}, Uception \cite{sanchesa2019cerebrovascular}, and RENet \cite{zhang2020cerebrovascular}. All deep learning methods we reproduced are trained until convergence by using the steps reported from their original publications. The proposed VOF-Net and VOF-Net without Oriented Filtering Module (backbone) were also tested. Fig. \ref{fig4} shows the results of each method on two MRA testing volumes and two CTA testing volumes. The second and fourth rows are the results of using Curved Planar Reformation (CPR) \cite{kanitsar2002cpr_10} technology to display one of the blood vessels. The images in the yellow boxes on the first and third rows are the close-ups of some vessels. Overall, all methods demonstrate a similar performance on main blood vessel branches. However, the segmentation results of our proposed method show better connectivity and more accurate boundary in detecting small vessels compared to other algorithms, as shown in the yellow rectangles and the CPR images.

Table 1 shows the performance of each method in all indicators, and the last column shows the amount of parameters of each method. As we can see, for all methods, the results on the CTA dataset generally outperform the MRA dataset on DSC, which may be attributed to the fact that the annotated vessels of the public MRA dataset are thin and includes some venous structures next to arterial vessels \cite{hilbert2020brave}. Another observation is that DSC score of backbone (VOF-Net without Oriented Filtering Module) is 2.11\% and 5.97\% lower than that of VOF-Net on MRA and CTA dataset, and it also has worse performance in other indicators. This indicates that embedding the prior knowledge of vessel structure by oriented filtering is effective for blood vessel segmentation. In addition, compared to REnet, our proposed method is slightly better, but with much fewer parameters. Overall, the proposed VOF-Net yields the best results in terms of all metrics. But our proposed system is a two-stage method, not end-to-end, which is a deficiency compared with other methods.

\section{Conclusion}

In this paper, we propose a deep neural network method, VOF-Net, to segment blood vessels. The network structure combines traditional blood vessel filtering algorithm and deep learning method. The orientation field of blood vessels is first estimated to guide oriented filtering. Then, the features of vessel structure extracted by oriented filtering are embedded into the vessel segmentation network, so as to improve the performance in vessel segmentation. Experimental results on datasets of CTA and MRA demonstrate the effectiveness of the proposed method. Although it was trained and evaluated on cerebrovascular segmentation, the
proposed method should be applicable to other tubular structures, such as coronary artery. It is our future work to extend the method to other tubular structures.

\bibliographystyle{splncs04_unsort}
\bibliography{refs}

\end{document}